\pdfoutput=1
\documentclass{article}
\usepackage{spconf,amsmath,graphicx}
\usepackage[hidelinks]{hyperref}
\usepackage{booktabs}
\usepackage{xcolor}
\usepackage{flushend} %

\title{REVERBERANT SPEECH CONDITIONING FOR MATERIAL-AWARE\\ ACOUSTIC PARAMETER MAPPING}

\name{Yi Wu$^{\star\dagger}$\thanks{This work was conducted while Yi Wu was an
  intern at Dolby Laboratories, Inc.} \qquad Olha Townsend$^{\star}$ \qquad
  Qianqian Fang$^{\star}$ \qquad Magdalena Fuentes$^{\dagger}$ \qquad
  You Zhang$^{\star}$}
\address{$^{\star}$Dolby Laboratories, Inc. \qquad $^{\dagger}$New York University}

\begin{document}
\ninept
\maketitle

\begin{abstract}
Dense maps of room acoustic parameters (T30, EDT, C50, DRR) describe how a room sounds at every position. Scene-wide estimation predicts such maps from a floor plan but requires a reference impulse response, which is hard to obtain. Our system instead receives a floor plan and easily recorded reverberant speech from one talker-microphone pair. We compare four speech-derived conditioning inputs: MFCCs, an embedding from the blind estimator BERP, BERP's scalar reverberation-time estimate, and an impulse response reconstructed by Rec-RIR. Our main benchmark is the MRAS dataset, where material assignments vary at fixed geometry. The reconstructed impulse response performs best among the four inputs, reducing T30 error from 31.5\% with the floor plan alone to 19.9\%, 83\% of the reduction achieved by the reference response. We find that adding speech mainly sets the overall level of the map, with little improvement in predicting how the acoustics vary across the room.
\end{abstract}
\begin{keywords}
Room acoustics, acoustic parameter estimation, blind estimation,
reverberant speech, acoustic parameter maps
\end{keywords}

\section{Introduction}
\label{sec:intro}

Room acoustics vary across positions within a room and across rooms, driven by geometry and material properties. To precisely describe a room's acoustics, one must measure one impulse response for every source-receiver pair. Room acoustic parameters compress each impulse response into a compact, perceptually relevant summary: reverberation time (T30), early decay time (EDT) and clarity (C50), defined in ISO~3382-1~\cite{iso3382}, plus the direct-to-reverberant ratio (DRR), per octave band. A room acoustic parameter map gives these parameters at every position in the room. Such maps support acoustic design, precomputed game and AR/VR rendering~\cite{raghuvanshi2014,raghuvanshi2018}, and parameter-conditioned impulse-response generation~\cite{paramrir2025}.

Producing room acoustic parameter maps typically requires simulation over a geometry with material properties, or impulse-response measurements at many positions~\cite{xrir2025}. Both are time consuming~\cite{raghuvanshi2014,tang2020} and unsuitable for a device in a room that was neither measured nor simulated~\cite{tang2020}, such as a smartphone, smart glasses or an XR headset.  In contrast, modern devices can reconstruct a room's floor plan, estimate their own position, and easily record speech in the room~\cite{gotz2026}. Such recordings are shaped by the room's acoustics and contain cues to its acoustic properties, such as how quickly sound decays at different frequencies.

\begin{figure}[t]
  \centering
  \includegraphics[width=0.86\linewidth]{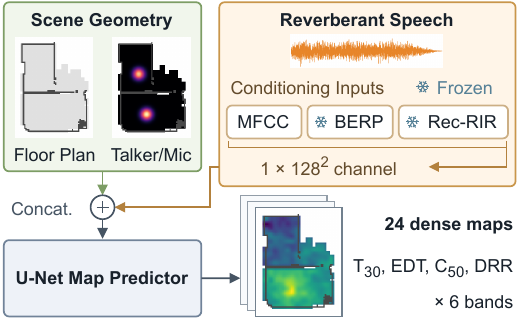}
  \caption{Illustration of our framework for acoustic parameter mapping. Reverberant speech is encoded into one of the speech-derived inputs, then concatenated with the floor map and fed into the map predictor, which predicts dense maps. Details in Sec.~\ref{ssec:model}.}
  \label{fig:overview}
  \vspace{-4pt}
\end{figure}

Two lines of work set the foundations for our investigation. One is \textit{blind estimation of room acoustic parameters from reverberant speech}. Such methods range from the reverberation-time and DRR estimators compared in the ACE challenge~\cite{ace2016} to later learning-based estimators~\cite{gotz2026, gamper2018,bryan2020,berp2025,ssbrpe2025,buddy2025} and blind reconstruction of the impulse response~\cite{fins2021,recrir2026}. With these methods, though speech is easy to capture, the parameter estimation is typically done for only one source-receiver position (i.e. talker-microphone) at a time. Some blind methods estimate acoustics beyond the recorded talker-microphone pair: Llu{\'i}s and Meyer-Kahlen~\cite{lluis2025} generate a spatial response at a queried source-receiver offset from a four-channel reverberant recording, and Tang et al.~\cite{tang2020} fit material absorption in a coarse room model from speech-estimated reverberation times and re-simulate the scene, but neither predicts dense maps of room acoustic parameters. The other is \textit{scene-wide estimation}~\cite{sae2025}, which predicts dense parameter maps, i.e., parameters for multiple source-receiver positions simultaneously. This is achieved by feeding a U-Net a floor plan with marked source and receiver positions, along with a reference impulse response from the scene, to predict dense per-band parameter maps. Related methods that predict impulse responses at unmeasured positions also rely on reference responses from the scene~\cite{xrir2025, naf2022,diffrir2024}. However, we argue that the reference impulse response from the target scene is typically hard to obtain. To our knowledge, no prior method directly predicts dense scene-wide maps of room acoustic parameters from a floor plan and speech.

In this work, we therefore investigate how effectively reverberant speech recorded at a talker-microphone pair can replace the impulse response of that pair, given a floor plan, for estimating room acoustic parameter maps. We first evaluate speech conditioning through two questions: (1) which conditioning input derived from speech most effectively replaces the reference response (Sec.~\ref{ssec:mras}); and (2) how well these systems generalize from simulation to real rooms without retraining (Sec.~\ref{ssec:rooms}). We also examine (3) how much of the reference-response benefit can be recovered with simpler acoustic inputs, and what speech changes in the predicted maps (Secs.~\ref{ssec:controls} and \ref{ssec:level}); and (4) whether a small calibration model can improve the predicted maps without retraining the U-Net (Sec.~\ref{ssec:calib}).

We contribute \textbf{(i) a method} for conditioning the U-Net map predictor of~\cite{sae2025} on reverberant speech, \textbf{(ii) a controlled comparison} of four conditioning inputs, and \textbf{(iii) an analysis} of what speech changes in the maps, which leads to a small calibration model.

\vspace{-1pt}

\section{Experimental Setup}
\label{sec:protocol}

\subsection{Methods}
\label{ssec:model}
\textbf{Model structure.} Our model is adapted from the U-Net of~\cite{sae2025}. As Fig.~\ref{fig:overview} shows, its input is the channel-wise concatenation of one of four conditioning inputs derived from speech and a rasterized floor plan (the floor map) with a scene mask and marked source (talker) and receiver (microphone) positions. It predicts maps of four parameters (T30, EDT, C50, DRR) in six octave bands, 24 maps in total. We call this the \textit{speech-conditioned} configuration.

\textbf{Speech-derived inputs.} All four inputs derive from the same utterance convolved with the talker-microphone impulse response supplied to the reference-response configuration. Each replaces the impulse-response spectrogram of~\cite{sae2025} as a $128\times128$ channel. We compare them to assess the benefits of learned features and explicit acoustic estimates. Mel-frequency cepstral coefficients (MFCCs) provide a spectral baseline. BERP~\cite{berp2025}, a blind estimator of room acoustic parameters, provides two conditioning inputs: its learned embedding and its scalar reverberation-time estimate. Rec-RIR~\cite{recrir2026} reconstructs the impulse response from speech. We call our four proposed systems \emph{spectral} (MFCCs), \emph{embedding} (BERP's embedding), \emph{scalar} (BERP's $T_{60}$ estimate), and \emph{reconstruction} (Rec-RIR).

\textbf{Implementation details.} The \emph{spectral} system uses 128-dimensional MFCC features computed with the same front end as BERP. We average-pool these features to 128 frames and standardize each coefficient. The \emph{embedding} system uses the 512-dimensional per-frame features of BERP's frozen encoder, pooled to 128 frames and mapped to the channel by a small learned adapter with a $1\times1$ convolution from 512 to 128 channels, followed by a GELU and a second $1\times1$ convolution from 128 to 128 channels (82k parameters). The \emph{scalar} system uses BERP's broadband reverberation-time estimate $T_{60}$ as a constant channel. The \emph{reconstruction} system uses Rec-RIR's blind reconstruction of the impulse response (a 2~s output at 16~kHz, resampled to 24~kHz), processed by the same spectrogram front end and 2~s response window as the reference response.

\textbf{Baseline configurations.} The original U-Net in~\cite{sae2025} concatenates the floor map with the magnitude mel spectrogram (128 mel frequency bins by 128 frames) of the room impulse response of the marked source-receiver pair; we use a 2~s window at 24~kHz to cover the longer MRAS responses. We use this U-Net as the \textit{reference-response} configuration, and we add a \textit{geometry-only} configuration in which the U-Net receives only the floor map.

\vspace{-1pt}
\subsection{Datasets}
\label{ssec:datasets}
\textbf{Simulated scenes.} MRAS~\cite{sae2025} provides impulse responses and acoustic parameter maps for procedurally generated multi-room floor plans built from shoe-box rooms. The acoustic simulations use the ray-tracing methods of SoundSpaces~\cite{soundspaces2020}. We use MRAS as our main benchmark because it varies the material assignment while keeping the floor plan fixed.  We use the preprocessed MRAS data released with~\cite{sae2025}, i.e. the authors' mono impulse responses at 24~kHz and their target maps, for 100 geometries from the grid-layout subset, each with up to five material assignments over the same floor plan. We split the geometries 60/10/30 into training, development and held-out test folds, with all material assignments of a geometry in the same fold. The test fold contains 21,600 talker-microphone pairs. The released MRAS target maps contain six octave-band channels labeled 125~Hz to 4~kHz. We verified from the released impulse responses and code that the corresponding band centers actually range from 31.5~Hz to 1~kHz, two octaves lower. We therefore relabel the channels accordingly.

\textbf{Speech and noise.} We convolve 8~s anechoic utterances of EARS~\cite{ears2024} talkers with the simulated impulse responses and add recorded indoor noise from one of three DEMAND~\cite{demand2013} environments at a fixed signal-to-noise ratio of 15~dB. Talkers are disjoint across the training, development and test folds, with twelve utterances at test. All material assignments of a talker-microphone pair use the same utterance and noise segment, so that changing the assignment changes only the impulse response.

\begin{table*}[t]
  \centering
  \caption{Speech conditioning on MRAS (test fold, three
  seeds). All results are from our own evaluation. Errors are aggregated over the six bands; T30/EDT errors are
  proportional. Gap closure (\%, Sec.~\ref{ssec:metrics}) for T30
  and EDT; Error increase under mismatch = paired increase (pp) in the
  same system's error when the observation (speech or reference response) comes from another material
  assignment of the same geometry (Sec.~\ref{ssec:mras}). Brackets: 95\% bootstrap intervals over
  the 30 test geometries (Sec.~\ref{ssec:metrics}); bold = best
  speech-conditioned system. For the T30 spatial analysis (one seed),
  $E_{\mathrm{pat}}$ (\%) is the error after optimal level correction, and $\rho$ is the
  Pearson correlation over in-room pixels (Sec.~\ref{ssec:level}).}
  \label{tab:ladder}
  \small
  \setlength{\tabcolsep}{1.6pt}
  \resizebox{\textwidth}{!}{
  \begin{tabular}{lcccccccccc}
    \toprule
    & \multicolumn{4}{c}{Error (matched observation) $\downarrow$} &
      \multicolumn{2}{c}{Gap closure $\uparrow$} &
      \multicolumn{2}{c}{Error increase under mismatch} &
      \multicolumn{2}{c}{Pattern (T30)} \\
    \cmidrule(lr){2-5} \cmidrule(lr){6-7} \cmidrule(lr){8-9} \cmidrule(l){10-11}
    Input & T30 (\%) & EDT (\%) & C50 (dB) & DRR (dB) & T30 (\%) & EDT (\%) & T30 (pp) & EDT (pp) & $E_{\mathrm{pat}}$ (\%) $\downarrow$ & $\rho$ $\uparrow$ \\
    \midrule
    Geometry-only & 31.5 & 30.2 & 2.68 & 1.65 & 0 & 0 & -- & -- & 12.8 & 0.61 \\
    \addlinespace
    \textit{scalar} (BERP $T_{60}$ estimate) & 25.2 & 25.2 & 2.29 & 1.49 & 45\,[38,51] & 55\,[48,61] & 10.3\,[8.5,12.2] & 10.0\,[8.5,11.6] & 12.5 & 0.63 \\
    \textit{embedding} (BERP) & 24.1 & 24.4 & 2.21 & 1.48 & 52\,[46,58] & 63\,[57,68] & 13.0\,[10.9,15.2] & 12.0\,[10.3,13.7] & 13.2 & 0.59 \\
    \textit{spectral} (MFCC) & 23.6 & 24.5 & 2.29 & 1.51 & 56\,[50,60] & 63\,[57,67] & 13.8\,[11.7,16.1] & 10.7\,[9.2,12.3] & 13.3 & 0.56 \\
    \textit{reconstruction} (Rec-RIR) & \textbf{19.9} & \textbf{21.8} & \textbf{2.08} & \textbf{1.43} & \textbf{83\,[78,86]} & \textbf{93\,[89,96]} & 18.6\,[15.8,21.5] & 15.2\,[13.1,17.3] & 12.8 & 0.60 \\
    \addlinespace
    Reference response & 17.5 & 21.1 & 1.98 & 1.40 & 100 & 100 & 23.4\,[19.9,27.0] & 17.3\,[14.9,19.9] & 12.7 & 0.60 \\
    \bottomrule
  \end{tabular}
  }
\end{table*}

\subsection{Training details}
\label{ssec:training}
We train the U-Net map predictor separately for the two configurations and the four systems, using the same MRAS training split and hyperparameters. Following \cite{sae2025}, we use a masked pixel-wise L1 loss and a batch size of 128. We train with the Ranger21 optimizer~\cite{ranger21} at a learning rate of $3\times10^{-4}$, validate every 1000 steps on talker positions withheld from the training geometries (about 30\% of them), and stop when the validation loss has not improved for five consecutive validations. Each is trained with three seeds. We use the released pretrained BERP\footnote{\url{https://github.com/Alizeded/BERP}} for the scalar and embedding systems and Rec-RIR\footnote{\url{https://github.com/Audio-WestlakeU/Rec-RIR}} for the reconstruction system. Both pretrained models remain frozen. For the \textit{embedding} system, we add a small adapter on BERP's frozen features, trained jointly with the U-Net.

\subsection{Metrics}
\label{ssec:metrics}
\textbf{Acoustic parameters and errors.} We evaluate map prediction accuracy for four room acoustic parameters~\cite{sae2025}, each in six octave bands. T30 describes how long reverberant sound persists, while EDT characterizes its initial decay; both are measured in seconds. C50 compares early with late sound energy, while DRR compares direct with reverberant sound energy; both ratios are expressed in dB. We focus on T30 and EDT because decay is sensitive to surface absorption not specified by the floor plan.  We compare predicted and ground-truth maps pixel by pixel, using proportional error $|\hat{y}-y|/y$ for T30 and EDT and absolute error in dB for C50 and DRR. We report T30 and EDT errors as percentages (\%) and differences between these errors in percentage points (pp).  Within each geometry, we average errors over in-room pixels, microphones, talkers, material assignments, seeds, and the six bands, in that order. We then average across geometries with equal weight. On the real rooms we average over receiver positions and seeds.

\textbf{Gap closure.} We compare each input's mean prediction error ($E_{\mathrm{sys}}$) with that of the geometry-only configuration ($E_{\mathrm{geo}}$). We introduce \emph{gap closure} to express the resulting error reduction as a fraction of the reduction achieved by the reference-response configuration (mean error $E_{\mathrm{ref}}$): $(E_{\mathrm{geo}}-E_{\mathrm{sys}})/ (E_{\mathrm{geo}}-E_{\mathrm{ref}})$, reported as a percentage. Higher is better: 0\% corresponds to the geometry-only configuration and 100\% to the reference-response configuration. Values above 100\% indicate lower error than the reference-response configuration. Because MRAS test errors are averaged over 30 geometries, we compute 95\% bootstrap intervals with 10,000 draws over geometries. We report standard deviations across training seeds separately to
quantify training variability.

\section{Evaluation of Speech Conditioning}
\label{sec:eval}

We first compare how effectively the four speech-derived conditioning inputs replace the reference response on MRAS (Sec.~\ref{ssec:mras}), then evaluate transfer to real rooms without retraining (Sec.~\ref{ssec:rooms}).

\subsection{Comparison of conditioning inputs on MRAS}
\label{ssec:mras}

\textbf{Comparison.} The matched observation columns (same material, same geometry) of Table~\ref{tab:ladder} show that all four speech-derived inputs improve T30 and EDT prediction over geometry-only. Rec-RIR gives the lowest errors among the speech inputs on all four parameters and recovers most of the reference-response benefit, closing 83\% of the T30 gap and 93\% of the EDT gap. Paired 95\% intervals support its lead over the other speech inputs on both parameters. BERP's scalar estimate already recovers about half of the benefit, while its embedding shows no consistent advantage over MFCCs. Over the four bands from 125\,Hz to 1\,kHz, where we expect the ray-tracing simulation behind MRAS~\cite{sae2025} to be reliable~\cite{kuttruff}, Rec-RIR still recovers most of the reference-response benefit. These results identify the reconstructed impulse response as the most effective speech-derived conditioning input tested on MRAS.

\textbf{Material-assignment mismatch.} To test whether the systems use material-dependent acoustic information, we replace the speech or reference response with one from each other material assignment of the same geometry, keeping the target map, utterance, noise, and positions fixed. For Rec-RIR, Table~\ref{tab:ladder} reports a T30 error increase of 18.6\,pp under mismatch: its matched error of 19.9\% rises to 38.5\%, above geometry-only's 31.5\%. All four speech-conditioned systems and the reference-response configuration then have higher mean T30 and EDT errors than geometry-only. The speech inputs' gains over geometry-only therefore depend on acoustic information from the target material assignment.

\textbf{Robustness.} Training variability is small for most speech inputs: across the three seeds, T30 and EDT error standard deviations are below 1\,pp except for MFCC (about 2\,pp).  The talker position is largely redundant: retraining the reconstruction system's map predictor with one position marker, interpreted as the device's position by acoustic reciprocity~\cite{kuttruff}, changes T30 error little ($-0.05$\,pp [$-0.18$, $+0.07$]) and increases EDT slightly ($+0.35$\,pp [$+0.18$, $+0.51$]). To test sensitivity to the speech corpus, we replace EARS with anechoic read speech from two speakers of the TSP Speech Database~\cite{tsp2002}. TSP is not listed in the documented training data for BERP~\cite{berp2025} or Rec-RIR~\cite{recrir2026}. On the development fold, without retraining, Rec-RIR's T30 and EDT errors increase by 0.28 and 0.43\,pp, respectively, but it remains the best speech input.

\subsection{Zero-shot transfer to real rooms}
\label{ssec:rooms}

To assess transfer from simulation to real rooms, we evaluate the MRAS-trained map predictors for all inputs in Table~\ref{tab:ladder} without retraining. We use the C4DM classroom~\cite{c4dm2010}, with one source and 130 receiver positions on a 0.5\,m lattice, and dEchorate~\cite{dechorate2021}, with four source positions and 30 microphones in ten surface configurations of one $6 \times 6 \times 2.4$\,m room. We construct floor plans from the published dimensions and convolve measured responses with the same anechoic utterances used for MRAS, adding the same noise at 15\,dB SNR. We compute target parameters following~\cite{sae2025} and evaluate from 250\,Hz to 1\,kHz. Both rooms use directional loudspeakers.

\textbf{C4DM classroom.} All four speech inputs reduce T30 error from geometry-only's 62.2\% to 18.8--26.5\% and EDT error from 63.9\% to 17.5--29.1\%. C50 and DRR errors also decrease. The reference-response configuration remains more accurate on both decay parameters, with 8.9\% T30 and 13.4\% EDT error. Speech conditioning therefore provides a substantial benefit in this room. The scalar system has the lowest mean T30 error among speech inputs, but the scalar, embedding, and reconstruction systems are within 0.7\,pp, so Rec-RIR's T30 advantage on MRAS does not carry over.

\textbf{dEchorate.} Transfer is more limited for T30: all four speech-conditioned systems predict values 1.9--2.5 times the measured values. At the input stage, BERP's median broadband estimate is 0.31\,s, above the measured median of 0.25\,s, whereas the broadband T30 median from Rec-RIR's reconstruction is closer at 0.23\,s. The reference-response configuration also overestimates T30 despite using measured impulse responses, pointing to a transfer limitation in the MRAS-trained map predictor. Shorter decay times may contribute: median broadband T30 is 0.25\,s in dEchorate and 0.76\,s in MRAS. To test whether predictions still use material-dependent acoustic information, we swap speech across surface configurations at fixed geometry and positions. This increases all four parameter errors for every speech-conditioned system, so sensitivity to surface-dependent acoustics persists despite the T30 overestimation.

Overall, zero-shot transfer is partly successful: speech conditioning substantially improves prediction in the C4DM classroom, but T30 remains overestimated in dEchorate.
\section{Analysis and Map Calibration}
\label{sec:analysis}

Section~\ref{sec:eval} shows that the reconstructed response recovers most of the reference-response benefit on MRAS, but leaves open how much can be recovered with simpler acoustic inputs (Sec.~\ref{ssec:controls}) and what speech changes in the maps (Sec.~\ref{ssec:level}). We investigate these questions, then test whether a calibration model can improve geometry-only maps without retraining (Sec.~\ref{ssec:calib}).

\begin{table}[t]
  \centering
  \caption{Gap closure (\%) on the MRAS test fold: upper, the U-Net trained per input (Sec.~\ref{ssec:controls}); lower, the calibration model (Sec.~\ref{ssec:calib}). 0 = geometry-only, 100 = the reference response; higher is better. Brackets: 95\% bootstrap intervals (Sec.~\ref{ssec:metrics}).}
  \label{tab:mechanism}
  \small
  \setlength{\tabcolsep}{2pt}
  \begin{tabular*}{\columnwidth}{@{\extracolsep{\fill}}lcc@{}}
    \toprule
    & \multicolumn{2}{c}{Gap closure $\uparrow$} \\
    \cmidrule(l){2-3}
    Input & T30 (\%) & EDT (\%) \\
    \midrule
    \multicolumn{3}{@{}l}{U-Net map predictor trained per input (three seeds):} \\
    BERP's $T_{60}$ estimate & 45\,[38,51] & 55\,[48,61] \\
    Rec-RIR's broadband T30 & 48\,[37,56] & 63\,[53,72] \\
    Rec-RIR reconstruction & 83\,[78,86] & 93\,[89,96] \\
    True broadband T30 & 63\,[54,71] & 66\,[56,75] \\
    True per-band T30 & 105\,[103,108] & 99\,[94,102] \\
    \midrule
    \multicolumn{3}{@{}l}{Geometry-only map calibrated (dev-fold fit, one seed):} \\
    BERP's $T_{60}$ estimate & 43\,[28,54] & 56\,[39,68] \\
    Rec-RIR broadband + per-band T30 & 43\,[26,57] & 59\,[42,73] \\
    True broadband T30 & 56\,[44,65] & 64\,[50,75] \\
    True per-band T30 & 86\,[80,91] & 76\,[61,87] \\
    \bottomrule
  \end{tabular*}
\end{table}

\subsection{Reverberation time as a conditioning input}
\label{ssec:controls}

BERP's scalar estimate already recovers about half of the reference-response benefit on the decay parameters (Sec.~\ref{ssec:mras}). We therefore test simpler acoustic inputs that replace the reference response with one broadband T30 value or six per-band values of that response. We compute T30 directly from the reference response to test whether blind-estimation errors fully explain the limited benefit of a broadband summary. We also extract broadband T30 from Rec-RIR's reconstruction to test how much of the reconstruction's benefit this single value retains. For each T30 input, we train a separate U-Net map predictor on MRAS, supplying each value as a constant channel.

In the upper block of Table~\ref{tab:mechanism}, six true per-band T30 values recover essentially all of the reference-response benefit: 105\% for T30 and 99\% for EDT. True broadband T30 recovers more than BERP's estimate but only part of the benefit. Thus, estimation error accounts for part of the scalar system's gap, but a single broadband value limits the benefit even when exact. Likewise, the broadband T30 of Rec-RIR's reconstruction recovers less than the reconstruction itself, suggesting useful conditioning beyond its broadband decay time.

\subsection{Global level and spatial pattern}
\label{ssec:level}

\begin{figure}[t!]
  \centering
  \includegraphics[width=0.9\linewidth]{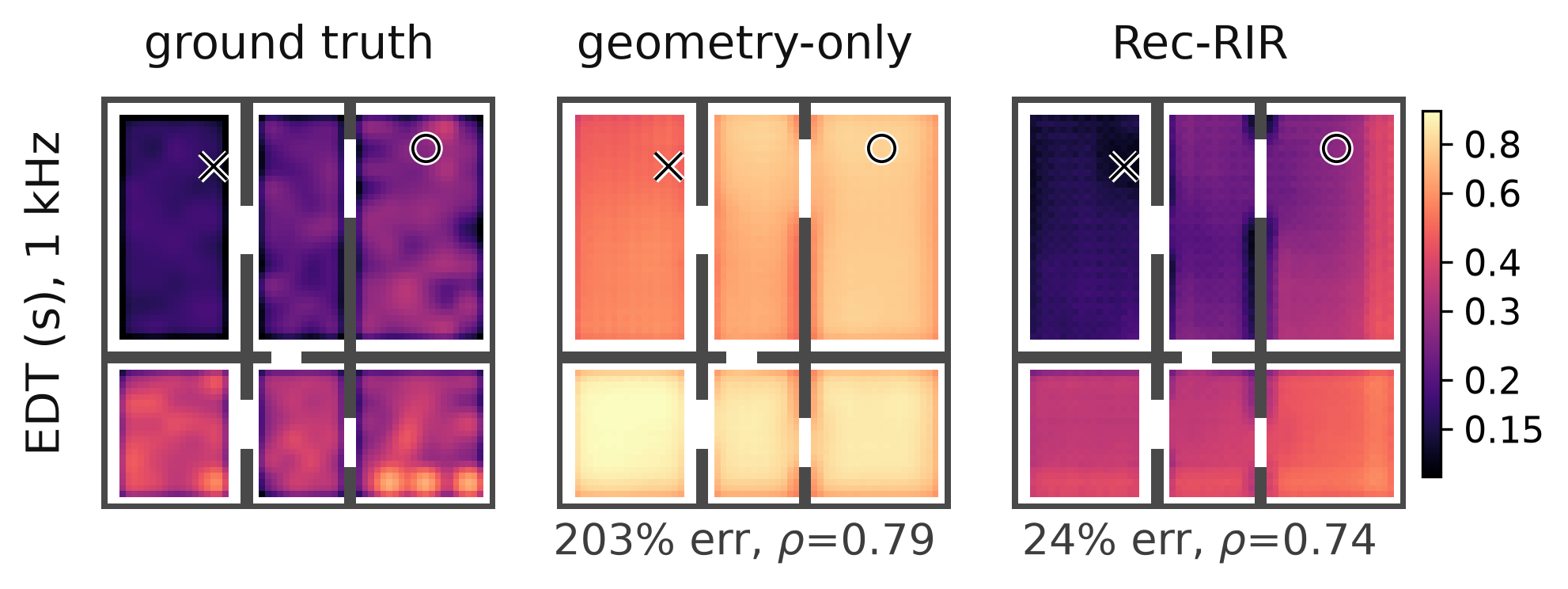}
  \vspace{-10pt}
  \caption{Test-fold example ($\times$ = talker, $\circ$ =
  observation microphone, in different rooms): EDT at 1\,kHz, ground
  truth, geometry-only, and Rec-RIR on a logarithmic color scale, with
  error and $\rho$. Rec-RIR reduces the error with
  similar $\rho$. Selected for a large improvement.}
  \label{fig:qual}
\end{figure}

To assess whether an acoustic input improves spatial accuracy beyond adjusting the map's overall level, we report $E_{\mathrm{pat}}$, the error remaining after optimal level correction. For each predicted map and band, we choose a positive multiplier (T30, EDT) or an additive dB offset (C50, DRR) that minimizes the error against the ground truth. This uniform correction preserves the map's spatial pattern, the relative variation across positions. We also compute the Pearson correlation $\rho$ between predicted and true values over valid in-room pixels to assess how well the prediction tracks the spatial variation.

In MRAS, materials change both the overall level and the spatial pattern of the maps. On the test fold, the same correction applied between the ground-truth T30 maps of two material assignments of the same geometry removes on average only 53\% [49,56] of their difference. We test which part the acoustic inputs recover. On one MRAS seed, the Pattern (T30) columns of Table~\ref{tab:ladder} show similar $E_{\mathrm{pat}}$ and $\rho$ for geometry-only, Rec-RIR, and the reference response. Every speech input has an $E_{\mathrm{pat}}$ within 0.5\,pp of geometry-only's. Across the tested inputs, $E_{\mathrm{pat}}$ also shows little improvement over geometry-only for EDT, C50, and DRR. Fig.~\ref{fig:qual} illustrates an EDT example in which Rec-RIR lowers the overall error with similar $\rho$. These results suggest that the gains from all tested acoustic inputs mainly reflect more accurate overall levels. Even the reference response yields little improvement in spatial accuracy with the current map predictor.

\subsection{Map calibration without retraining}
\label{ssec:calib}
Since speech mainly improves the map's overall levels (Sec.~\ref{ssec:level}), we test whether a small calibration model can recover these gains without retraining the geometry-only predictor. A network with two hidden layers of 64 units (about $10^4$ parameters) predicts one correction per parameter and band: a multiplier for T30 and EDT or a dB offset for C50 and DRR. Inputs are reverberation-time values, the in-room pixel count, and each geometry-only map's in-room mean and values at the talker and microphone pixels. We fit the calibration models on the development fold and evaluate them on the test fold using maps from one seed of the geometry-only predictor.

With BERP's $T_{60}$ estimate, the calibration model achieves similar T30 and EDT gap closure to the scalar system (Table~\ref{tab:mechanism}, lower and upper blocks). For C50 and DRR, it recovers about half the scalar system's benefit on the same seed. With true per-band T30 values, the calibration model recovers less of the reference-response benefit on both decay parameters than the U-Net trained with the same acoustic values. Thus, although the tested calibration model does not reproduce all gains from training the U-Net with acoustic inputs, it retains much of the scalar system's decay benefit without retraining the geometry-only predictor.

\section{Conclusion}
\label{sec:conclusion}

Reverberant speech carries material-dependent information, making room acoustic parameter maps predicted from a floor plan material-aware. The reconstructed impulse response is the best speech input on MRAS. Speech mainly sets the overall level of the map, with little improvement in spatial accuracy. A small calibration model recovers much of the scalar system's decay benefit without retraining the geometry-only predictor. Future work could test speech from several positions or furnished 3D geometry for the spatial pattern.

\clearpage
\bibliographystyle{IEEEbib}
\bibliography{refs}

\end{document}